\documentclass[preprint]{aastex}
\usepackage{natbib,psfig,emulateapj5}

\newcommand\rosat{{\sl ROSAT} }

\newcommand\ergs{{erg s$^{-1}$\thinspace}}

\newcommand\tx{\hbox{{$T_{\rm x}$}}}
\newcommand\lx{\hbox{{$L_{\rm x}$}}}

\begin{document} 

\title{On The Origin of Radio Halos in Galaxy Clusters}

\author{David A. Buote}

\affil{Department of Physics and Astronomy, University of California
at Irvine, 4129 Frederick Reines Hall,\\ Irvine, CA 92697-4575;
buote@uci.edu} 

\slugcomment{Accepted for Publication in The Astrophysical Journal Letters}

\begin{abstract}
Previously it has been recognized that radio halos in galaxy clusters
are preferentially associated with merging systems as indicated by
substructure in the X-ray images and temperature maps. Since, however,
many clusters without radio halos also possess substructure, the role
of mergers in the formation of radio halos has remained unclear.  By
using power ratios to relate gravitational potential fluctuations to
substructure in X-ray images, we provide the first quantitative
comparison of the dynamical states of clusters possessing radio halos.
A correlation between the 1.4 GHz power ($P_{1.4}$) of the radio halo
(or relic) and the magnitude of the dipole power ratio ($P_1/P_0$) is
discovered such that approximately $P_{1.4}\propto P_1/P_0$; i.e., the
strongest radio halos appear only in those clusters currently
experiencing the largest departures from a virialized state.  From
additional consideration of a small number of highly disturbed
clusters without radio halos detected at 1.4 GHz, and recalling that
radio halos are more common in clusters with high X-ray luminosity
(Giovannini, Tordi, \& Feretti), we argue that radio halos form
preferentially in massive ($\lx\ga 0.5 \times 10^{45}$
\ergs) clusters experiencing violent mergers ($P_1/P_0\ga 0.5 \times
10^{-4}$) that have seriously disrupted the cluster core. The
association of radio halos with massive, large-$P_1/P_0$,
core-disrupted clusters is able to account for both the vital role of
mergers in accelerating the relativistic particles responsible for the
radio emission as well as the rare occurrence of radio halos in
cluster samples.
\end{abstract}


\section{Introduction}
\label{intro}

Diffuse radio emission that cannot be attributed only to individual
galaxies in a galaxy cluster is termed a {\it radio halo} if the
emission is centrally located or a {\it radio relic} if it lies
substantially away from the (X-ray) cluster center (for reviews see,
e.g., Feretti 2000; Sarazin 2000). Radio halos typically extend to 1
Mpc scales and are characterized by a steep radio spectrum consistent
with a synchrotron origin.  Until recently radio halos were known to
exist in only a handful of galaxy clusters with Coma being the
best-studied example (e.g., Giovannini et al. 1993; Deiss et
al. 1997). With the completion of the NRAO VLA Sky Survey
\citep{vlass} the number of candidate radio halos has risen to
approximately 20 (Giovannini, Tordi, \& Feretti 1999; Giovannini \&
Feretti 2000; Liang et al. 2000). However, these radio halos still
represent only $\sim 10\%$ of the cluster populations studied
indicating that they are indeed a rare phenomenon.

Important progress in our understanding of the formation of radio
halos has been made recently. First, \citet{govoni} have compared the
point-to-point spatial distribution of the radio and X-ray emission in
four clusters and have found a linear relationship in two cases and a
nearly linear relationship in the other two. The similarity of the
radio and X-ray morphologies suggests a direct connection between the
thermal X-ray plasma and the non-thermal radio plasma. Second,
\citet{sergio} and \citet{liang} have discovered a correlation between
radio power $P_{1.4}$ (at 1.4 GHz rest frame) and X-ray temperature
(\tx) such that $P_{1.4}$ increases for larger \tx\, in their sample
of 10 of the most securely detected radio halos. \citet{liang} suggest
that the $P_{1.4}-\tx$ correlation also indicates a direct connection
between the radio and X-ray plasmas.

Although there is mounting evidence that the thermal X-ray and
non-thermal radio emission are directly related, the source of the
relativistic particles giving rise to the non-thermal emission, as
well as the question of the rarity of radio halos, remains
unexplained. Perhaps the most favored mechanism to accelerate
relativistic electrons in clusters is that of mergers (e.g., Tribble
1993) owing to the considerable amount of energy available during a
merger ($\sim 10^{64}$ erg). The details of this process, however,
remain controversial because of the difficulty in directly
accelerating the thermal electrons to relativistic energies (e.g.,
Tribble 1993; Sarazin 2000; Brunetti et al. 2001; Blasi 2001).  In
fact, owing to this difficulty it is often assumed that a reservoir of
relativistic particles is established at some time in the past
evolution of the cluster with the current merger merely serving to
re-accelerate relativistic particles from this reservoir. In this case
it is unclear whether the current or the past dynamical state of the
cluster is the primary factor in the creation of a radio halo.

X-ray observations provide circumstantial evidence for a connection
between cluster merging and radio halos (see Feretti 2000 and
references therein) because, in particular, radio halos are only found
in clusters possessing X-ray substructure and weak (or non-existent)
cooling flows. However, it has been argued (e.g., Giovannini \&
Feretti 2000; Liang et al. 2000; Feretti 2000) that merging cannot be
solely responsible for the formation of radio halos because at least
50\% of clusters show evidence for X-ray substructure (Jones \& Forman
1999) whereas only $\sim 10\%$ possess radio halos. (Note X-ray and
optical substructures are well-correlated -- Kolokotronis et
al. 2001.)

Unfortunately, it is difficult to interpret the importance of merging
using the observed frequency of substructure as it does not itself
quantify the deviation of an individual cluster from a virialized
state. And the shocks that could be responsible for particle
acceleration will be proportionally stronger in clusters (of the same
mass) with the largest departures from a virialized state.  To measure
the dynamical states of clusters from X-ray images it is necessary to
quantify the cluster morphologies using statistics such as the
center-shift (Mohr, Fabricant, \& Geller 1993; Mohr et al. 1995) and
the power ratios (Buote \& Tsai 1995,1996; Buote 1998).

In this Letter we provide the first quantitative comparison of the
dynamical states of clusters possessing detected radio halos. Using
the power ratios computed previously from \rosat X-ray images we find
a correlation between the power of the radio halo and the magnitude of
the dipole power ratios; i.e., the strongest radio halos appear only
in those clusters currently experiencing the largest departures from a
virialized state.  We argue that this correlation confirms the vital
role of mergers in accelerating the relativistic particles responsible
for the radio emission and also explains the rarity of radio halos.

\section{Radio Power and Cluster Dynamical States}
\label{radio}

\begin{figure*}[t]
\parbox{0.49\textwidth}{
\centerline{\psfig{figure=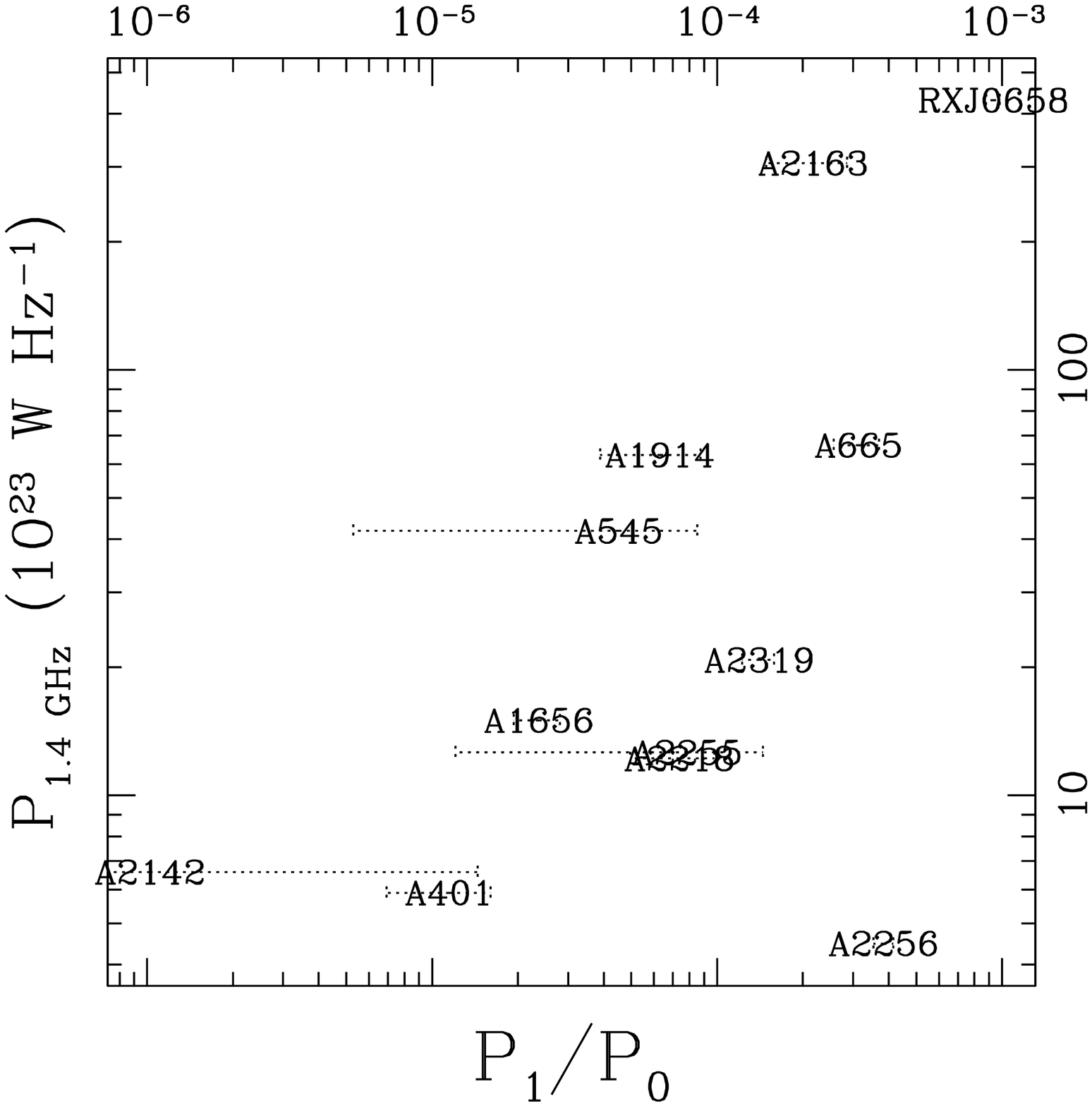,height=0.35\textheight}}}
\parbox{0.49\textwidth}{
\centerline{\psfig{figure=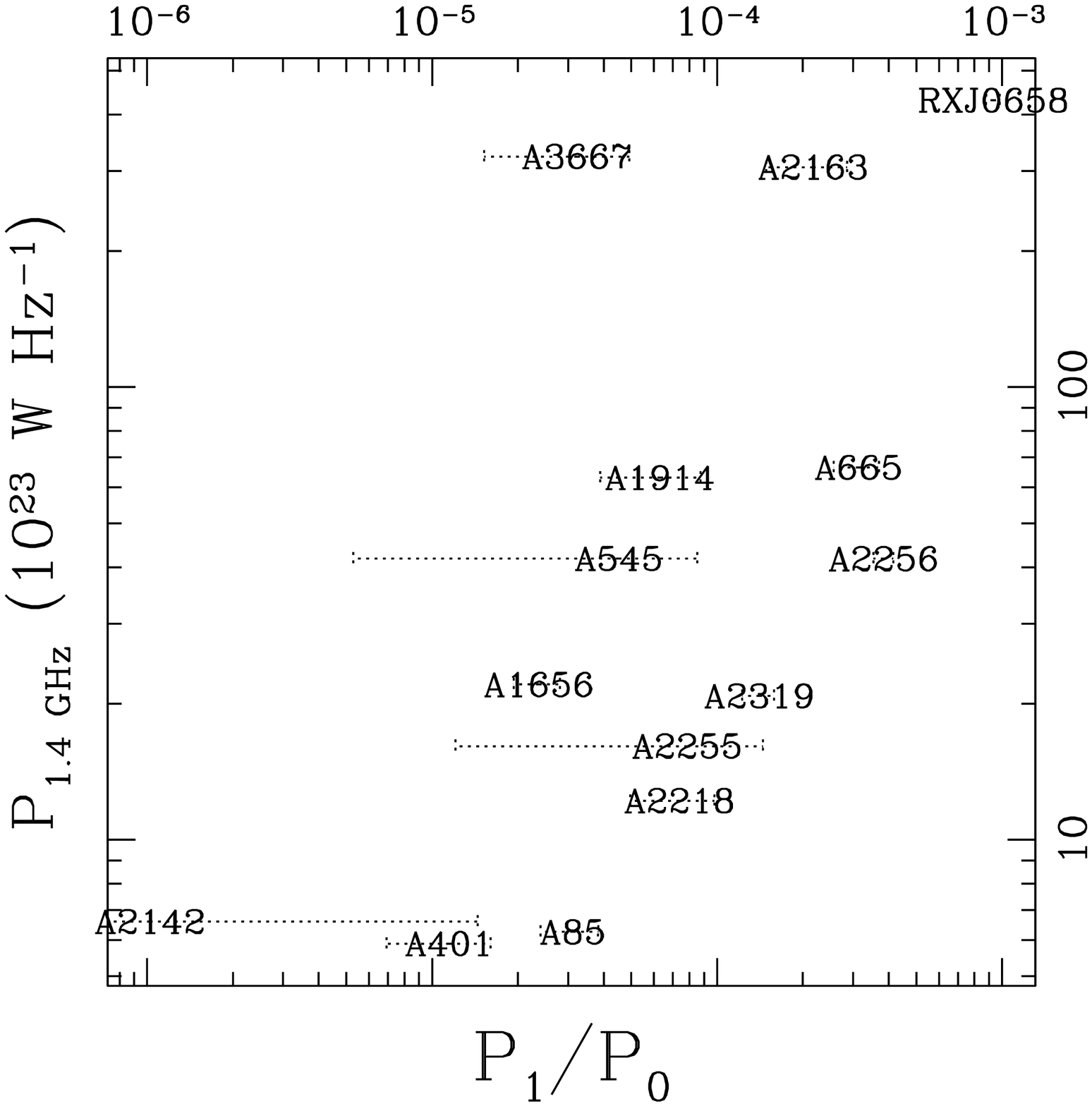,height=0.35\textheight}}}
\caption{\label{fig.radio} Radio power ($P_{1.4}$ -- 1.4 GHz rest
frame) versus dipole power ratio ($P_1/P_0$) where $P_{1.4}$ includes
emission from (Left) only radio halos and (Right) the total diffuse
emission from halos and relic sources. The power ratios are computed
within a 0.5 Mpc aperture centered on the X-ray emission peak with
estimated $1\sigma$ errors shown. (Uncertainties on $P_{1.4}$ are
believed to be $\la 10\%$ and are not shown.) }
\end{figure*}      

We compiled a sample of 14 clusters selected primarily from the
catalogs of radio halos and relics of \citet{feretti00},
\citet{gtf}, and \citet{gf} that also have power ratios measured by
BT96. (Note that the radio powers estimated from the NVSS survey may
be underestimated because of the high noise level -- e.g., Giovannini
et al. 1999.) Also included is the cluster RXJ0658-5557 (1E0657-56)
that has been reported to possess the most powerful radio halo to date
\citep{liang} even though it was not analyzed by BT96.  We obtained a
deep HRI exposure (58ks) of RXJ0658 from the \rosat public data
archive and computed power ratios in a manner similar to that
described by BT96.

It can be shown that the power ratios are a direct measure of the
dynamical state of a cluster modulo projection effects \citep{b98}.
Briefly, each $P_m$ represents the square of the $m$th multipole of
the two-dimensional pseudo-potential generated by the X-ray surface
brightness evaluated over a circular aperture. The aperture is
positioned at the peak of the X-ray image to compute the dipole power
ratio, $P_1/P_0$, but is located at the center of mass (surface
brightness) to compute the higher order moments (see BT96).  Large
departures from a virialized state are then indicated by large power
ratios for the lowest-order multipoles since they contribute most to
the potential.

In Figure \ref{fig.radio} we plot radio power versus $P_1/P_0$
evaluated over a 0.5 Mpc aperture radius. (We assume $H_0=80$ km
s$^{-1}$ Mpc$^{-1}$ and $q_0=0$ following BT96.)  Although the sample
is small, a clear trend is observed: radio power tends to increase
with increasing $P_1/P_0$, or the magnitude of the deviation from a
virialized state, such that approximately $P_{1.4}\propto P_1/P_0$.
Moreover, all radio-halo clusters except A2142 have $P_1/P_0\ga
10^{-5}$ whereas essentially all massive cooling flow clusters
populate $P_1/P_0\la 10^{-5}$ (see BT96). Since most of the radio
halos in Figure \ref{fig.radio} have $P_1/P_0$ values indicative of
unrelaxed systems, this confirms previous circumstantial evidence that
radio halos preferentially exist in merging systems without massive
cooling flows (e.g., Feretti 2000).

When considering only the emission from radio halos the cluster A2256
appears to be an outlier; i.e., it possesses a large value of
$P_1/P_0$ for a weak radio halo. However, when the emission from both
radio halos and relics are considered A2256 fits in much better (right
panel in Figure \ref{fig.radio}). Another cluster, A85, that only
possess a radio relic, also supports the trend of larger $P_{1.4}$ for
larger $P_1/P_0$ for both halos and relics.

The relic-only cluster A3667 appears to be a genuine outlier in that
it has a relatively small value of $P_1/P_0$ for its large radio
power. In principle this deviation could be attributed to a projection
effect (i.e., smearing of substructure).  But the relic source of
A3667 is offset $\sim 1$ Mpc from the X-ray center, and $P_1/P_0$ is
twice as large in the 1 Mpc aperture indicating a substantial
large-scale dynamical disturbance. In fact, the general
$P_{1.4}-P_1/P_0$ trend discussed above applies also when $P_1/P_0$ is
computed within a 1 Mpc aperture, but in this case A3667 has a large
value of $P_1/P_0$ similar to (for example) A2163 consistent with the
trend. Consequently the ``anomalous'' position of A3667 can be
attributed to $P_1/P_0$ computed within the 0.5 Mpc aperture not being
a sensitive enough indicator of the dynamical disturbance associated
with the relic on $\sim 1$ Mpc scales. An appropriately weighted
average of $P_1/P_0$ over an entire cluster is probably required to
obtain a fully consistent representation of halos and relics in the
$P_{1.4}-P_1/P_0$ plane.

We mention that for the power ratios computed within apertures located
at the center of mass (surface brightness) only the odd power ratio,
$P_3/P_0$, clearly displays a trend similar to $P_1/P_0$. However, the
statistical uncertainties for this higher order moment are large, and
the correlation is only observed for systems having $P_{1.4} >
10^{24}$ W Hz$^{-1}$. The even power ratios (e.g., $P_2/P_0$) do not
show a correlation with $P_{1.4}$. Apparently only dynamical
disturbances that contribute primarily to the odd power ratios
correlate with the power of radio halos and relics.

\section{Discussion}
\label{disc}

\subsection{Radio Halo and Relic Formation}

Previously it has been recognized that radio halos and relics tend to
be associated with mergers, but since a higher percentage of clusters
without radio halos show evidence for substructure the importance of
merging in the formation of radio halos has remained unclear (e.g.,
Feretti 2000).  The $P_{1.4}-P_1/P_0$ correlation (Fig
\ref{fig.radio}) not only confirms previous circumstantial evidence
relating the presence of radio halos to mergers but, more importantly,
establishes for the first time a quantitative relationship between the
``strength'' of radio halos and relics ($P_{1.4}$) and the
``strength'' of mergers ($P_1/P_0$). Moreover, in the
$P_{1.4}-P_1/P_0$ plane both radio halos and relics may be described
consistently which provides new evidence that both halos and relics
are formed via mergers.  The $P_{1.4}-P_1/P_0$ correlation supports
the idea that shocks in the X-ray gas generated by mergers of
subclusters accelerate (or re-accelerate) the relativistic particles
responsible for the radio emission.

\subsection{Implications of Outliers}
\label{outliers}

Most of the clusters studied by BT96 do not possess radio halos or
relics. If we consider the brightest $\sim 30$ clusters for which the
sample of BT96 is more than 50\% complete (X-ray selected) most of the
clusters without radio halos or relics have $P_1/P_0\la 10^{-5}$
placing them in the lower left portion of Figure
\ref{fig.radio}. These clusters are therefore approximately relaxed
systems and, in accordance with the formation scenario discussed
previously, do not have powerful radio halos or relics.

However, three bright clusters (A754, A3266, Cygnus-A) exist that are
highly morphologically disturbed ($P_1/P_0 \ge 10^{-4}$) yet have no
(or only weakly) detected emission from a radio halo at 1.4 GHz. Thus
each would lie in the bottom right portion of Figure \ref{fig.radio}
as significant outliers in the $P_{1.4}-P_1/P_0$ correlation.  It is
possible that powerful radio halos for these systems have not been
detected at 1.4 GHz owing to their steep spectra; e.g., after our
paper was submitted we became aware of a paper by \citet{kassim} which
presents evidence for a very powerful radio halo in A754 at 330 MHz.
Whether halos in these or other clusters are detected at other radio
frequencies is an important subject for future studies. For the
remainder of this paper we confine our discussion to studies at 1.4
GHz.

These clusters do have strong radio emission from either a central
source (Cygnus-A) or collectively from several point sources (A754 and
A3266) which could be related to their current dynamical state.  These
clusters have similar structure in their X-ray temperature
distribution where relatively cool gas exists within the central few
hundred kpc and hotter gas, consistent with shock-heating, is located
at larger radii (e.g., Henry \& Briel 1995; Henriksen \& Markevitch
1996; Markevitch et al. 1999,2001; Henriksen et al. 2000; Sarazin
2000). The temperature maps of these systems imply mergers that have
not disrupted the cores, and detailed hydrodynamical models confirm
that the mergers in these systems are off-axis and must be in the very
earliest stages (e.g., Roettiger et al. 1998; Flores et al. 2000;
Roettiger \& Flores 2000).

The temperature structure of these deviant clusters is similar to that
of A3667 (e.g, Vikhlinin et al. 2000) which also has no detected radio
halo. As discussed previously, this system has a large-scale dynamical
disturbance with a large value of $P_1/P_0\approx 10^{-4}$ within the
1 Mpc aperture similar to those clusters with the most powerful
halos. In contrast, A2256, the most deviant cluster in Figure
\ref{fig.radio} when considering only the emission from radio halos,
does have a measured weak radio halo.  Unlike the other clusters
described in this section, the X-ray temperature map of A2256 (e.g.,
Sun et al. 2001) indicates a more advanced merger that has begun to
disrupt the core.

{\it The properties of these deviant clusters suggest that radio halos
form only when a sufficiently large dynamical disturbance has
proceeded fully into the core of a cluster.} The formation of halos
and relics also appears to be related since the relic sources (most
notably A2256) are consistent with the $P_{1.4}\propto P_1/P_0$ trend
when both halo and relic emission are included. Further study is
needed to establish the existence of a direct link between halos and
relics, especially to ascertain if peripheral relics are formed
preferentially at early times during mergers.

Finally, the faintest cluster studied by BT96, A514, has the largest
power ratios but does not possess a radio halo. This cluster consists
of several small clumps embedded in a diffuse halo of X-ray emission
(e.g., Figure 5 in Buote \& Tsai 1995). The lack of a radio halo could
arise because A514 is apparently in the earliest formation stages and
perhaps has not had enough time to generate a reservoir of
relativistic particles for re-acceleration. Alternatively, the low
mass of this cluster may indicate that insufficient energy is
available to accelerate particles to the speeds required for
synchrotron emission.
 
\subsection{The Importance of the Mass of the Cluster}

Although $P_{1.4}\propto P_1/P_0$ (Figure \ref{fig.radio}) holds
approximately for systems with radio halos (and no relics) there is
considerable scatter for a given value of $P_1/P_0$.  For example,
A665 and A2163 have similar values of $P_1/P_0$ but differ by a factor
of $\approx 5$ in radio power.  It is possible that projection effects
account for the similar values of $P_1/P_0$ for A665 and A2163 which
will become apparent as the sample of radio halos with computed values
of $P_1/P_0$ increases. Or perhaps the dynamical states of these
clusters could be distinguished by using a radially averaged value of
$P_1/P_0$ (\S \ref{radio}). However, this large difference in radio
power could imply the existence of another physical parameter
fundamental to the formation of radio halos.

The mass of the cluster is a logical candidate for a fundamental
parameter since the energy available to accelerate relativistic
particles during a merger scales as $\sim M^2$. The positive
correlations of $P_{1.4}$ with \lx\, and \tx\, discovered by
\citet{gtf}, \citet{sergio}, and \citet{liang} provide strong evidence
for the influence of the cluster mass on the power of radio
halos. This mass dependence would explain the lack of a radio halo in
the highly disturbed low-luminosity cluster A514 discussed above.

It is possible that the scatter noted above for the relation
$P_{1.4}\propto P_1/P_0$ could be reduced if cluster masses are
considered.  For example, if $M\sim \tx^{3/2}$ as appropriate for pure
gravitational infall, then using the temperatures of \citet{daw00} we
obtain values of $(P_1/P_0)\tx^{3/2}$ of $7.1\times 10^{-4}$ for A665
and $10.5\times 10^{-4}$ for A2163; i.e., the cluster with larger
$P_{1.4}$ now also has larger $(P_1/P_0)\tx^{3/2}$. Observations of a
large sample of clusters will be required to determine whether the
mass, projection effects, the method of computing $P_1/P_0$, or the
frequency used to evaluate the radio power can reduce the scatter in
Figure \ref{fig.radio}.

\subsection{Rarity of Radio Halos and Relics}

The infrequent occurrence of radio halos in X-ray cluster samples can
be understood when both the dynamical state and mass of the cluster
are considered. In their study of an X-ray flux limited sample of 205
clusters \citet{gtf} detect radio halos in only $<5\%$ of clusters
with $\lx<0.5\times 10^{45}$ \ergs but in $\approx 30\%$ of clusters
with $\lx>1.0\times 10^{45}$ \ergs ($H_0=50$ km s$^{-1}$ Mpc$^{-1}$
and $q_0=0.5$). Hence, the need for sufficient $\lx$, therefore mass,
can explain the rarity of radio halos in lower mass clusters.

For the most massive and X-ray luminous systems the relative frequency
($<50\%$) of radio halos cannot be explained by considering only the
mass or, equivalently, only the X-ray temperature. For example, both
A665 and A2029 have $\tx\approx 8$ keV(e.g., White 2000) but only A665
has a powerful radio halo. However, A665 is currently experiencing a
violent merger (large $P_1/P_0$) whereas A2029 is apparently a nearly
relaxed system (small $P_1/P_0$). In fact, of the brightest $\sim 30$
clusters in the sample of BT96 virtually all clusters with
$P_1/P_0>0.5\times 10^{-4}$ either possess radio halos (relics) or
suggest an early merger that has not fully disrupted the cluster core
(see \S \ref{outliers}).  One can select just for the radio halos by
eliminating clusters that demonstrate characteristic X-ray temperature
structure (see \S \ref{outliers}) and perhaps also those with an
increasing $P_1/P_0$ radial profile (e.g., the outliers A3266 and
Cygnus-A -- see BT96).

Thus, for massive clusters the occurrence of radio halos may be
explained by the frequency of clusters currently experiencing violent,
core-disrupting, mergers. On average $P_1/P_0$ is expected to increase
with increasing redshift owing to the higher incidence of merging
\citep{b98} which would lead to a higher incidence of radio
halos. However, on average cluster masses are lower at earlier times
implying a lower incidence of radio halos. Each of these factors is
dependent on the assumed cosmology, and future theoretical work is
therefore required to establish whether the abundance of radio halos
(1) increases or decreases with redshift, and (2) provides an
interesting test of cosmological models.

\acknowledgements

I thank the anonymous referee for many helpful comments, especially
with regards to the published radio data.


\begin{thebibliography}{}
\bibitem[Blasi(2001)]{blasi}
Blasi, P. 2001, Astroparticle Physics, in press (astro-ph/0008113)
\bibitem[Brunetti et al.(2001)]{brun}
Brunetti, G., Setti, G., Feretti, L., \& Giovannini, G. 2001, \mnras,
320, 365
\bibitem[Buote(1998)]{b98}
Buote, D. A 1998, \mnras, 293, 381
\bibitem[Buote \& Tsai(1995)]{bt95}
Buote, D. A., \& Tsai, J. C. 1995, \apj, 452, 522
\bibitem[Buote \& Tsai(1996)]{bt96}
Buote, D. A., \& Tsai, J. C. 1996, \apj, 458, 27
\bibitem[Colafrancesco(1999)]{sergio}
Colafrancesco, S. 1999, in Diffuse Thermal and Relativistic Plasma in
Galaxy Clusters, eds. H. Boehringer, L. Feretti, P. Schuecker, MPE
Report 271, 269 (astro-ph/9907329) 
\bibitem[Condon et al.(1998)]{vlass}
Condon, J. J., Cotton, W. D., Greisen, W. W., Yin, Q. F., Perley,
R. A., Taylor, G. B., \& Broderick, J. J., 1998, \aj, 115, 1693
\bibitem[Diess et al.(1997)]{diess}
Deiss, B. M., Reich, W., Lesh, H., \& Wielebinski, R. 1997, \aap, 321,
55
\bibitem[Feretti(2000)]{feretti00}
Feretti, L. 2000, The Universe at Low Radio Frequencies, IAU 199, Pune
(India), in press (astro-ph/0006379)
\bibitem[Flores et al.(2000)]{flores}
Flores, R. A., Quintanan, H., \& M. J. Way 2000, \apj, 532, 206
\bibitem[Giovannini et al.(1993)]{gio93}
Giovannini, G., Feretti, L., Venturi, T., Kim, K.-T., \& Kronberg,
P. P. 1993, ApJ, 406, 399
\bibitem[Giovannini \& Feretti(2000)]{gf}
Giovannini, G., \& Feretti, L. 2000, New Astronomy, 5, 355
\bibitem[Giovannini et al.(1999)]{gtf}
Giovannini, G., Tordi, M., \& Feretti, L. 1999, New Astronomy, 4, 141
\bibitem[Govoni et al.(2001)]{govoni}
Govoni, F., Ensslin, T. A., Feretti, L., \& Giovannini, G. 2001, \aap,
in press (astro-ph/0101418)
\bibitem[Henriksen \& Markevitch(1996)]{hm96}
Henriksen, M. J., \& Markevitch M. 1996, \apjl, 466, L79
\bibitem[Henriksen et al.(2000)]{hdd}
Henriksen, M. J., Donnelly, R. H., \& Davis, D. S. 2000, \apj, 529,
692 
\bibitem[Henry \& Briel(1995)]{hb95}
Henry, J. P., \& Briel U. G. 1995, \apjl, 443, L9
\bibitem[Jones \& Forman(1999)]{jf99}
Jones, C., \& Forman, W. 1999, \apj, 511, 65
\bibitem[Kassim et al.(2001)]{kassim}
Kassim, N., Clarke, T. E., Ensslin, T. A.,Cohen, A. S., \& Neumann
2001, \apj, in press (astro-ph/0103492) 
\bibitem[Kolokotronis et al.(2001)]{kbpg}
Kolokotronis, V., Basilakos, S., Plionis, M., Georgantopoulos,
I. 2001, \mnras, 320, 49
\bibitem[Liang et al.(2000)]{liang}
Liang, H., Hunstead, R. W., Birkinshaw, M., \& Andreani, P. 2000,
\apj, 544, 686
\bibitem[Markevitch et al.(1999)]{msv}
Markevitch, M., Sarazin, C. L., Vikhlinin, A. 1999, \apj, 521, 526
\bibitem[Markevitch et al.(2001)]{m01}
Markevitch, M., Vikhlinin, A., Mazzotta P., Van Speybroeck L. 2001,  
X-ray astronomy 2000, eds R. Giacconi, L. Stella, S. Serio, ASP
Conf. Series, (astro-ph/0012215)
\bibitem[Mohr et al.(1993)]{mfg}
Mohr, J. J., Fabricant, D. G., \& Geller, M. J. 1993, \apj, 413, 492 
\bibitem[Mohr et al.(1995)]{mefg}
Mohr, J. J., Evrard, A. E., Fabricant, D. G., \& Geller, M. J. 1995,
\apj, 447, 8 
\bibitem[Roettiger \& Flores(2000)]{rf00}
Roettiger, K., \& Flores, R. 2000, \apj, 538, 92
\bibitem[Roettiger et al.(2000)]{rbs}
Roettiger, K., Burns, J. O., \& Stone, J. M. 2000, \apj, 518, 603
\bibitem[Sarazin(2000)]{sar00}
Sarazin, C. L. 2000, Constructing the Universe with Clusters of
Galaxies, eds F. Durret \& D. Gerbal, in press (astro-ph/0000904)
\bibitem[Sun et al.(2001)]{sun01}
Sun, M., Murray, S. S., Markevitch, M., \& Vikhlinin, A. 2001, \apj,
submitted (astro-ph/0103103)
\bibitem[Tribble(1993)]{tribble}
Tribble, P. C. 1993, \mnras, 263, 31
\bibitem[Vikhlinin et al.(2000)]{vmm}
Vikhlinin A., Markevitch, M., \& Murray, S. 2000, \apj, submitted
(astro-ph/0008496) 
\bibitem[White(2000)]{daw00}
White, D. A., 2000, \mnras, 312, 663 
\end{thebibliography}
\end{document}